\documentclass[pre,preprintnumbers,amsmath,amssymb,floatfix,12pt,authoryear]{revtex4}

\usepackage{graphics}
\usepackage{amsmath}
\usepackage{graphicx,color}
\usepackage[latin1]{inputenc}
\usepackage[T1]{fontenc}
\usepackage{ae,aecompl}

\begin{document}

\begin{titlepage}
\vskip .27in
\begin{center}
{\large \bf     Gallavotti-Cohen
theorem,  Chaotic Hypothesis and the zero-noise limit. }

\vskip .27in

{Jorge Kurchan}
\vskip .2in {\it PMMH-ESPCI, CNRS UMR 7636, 10 rue Vauquelin,
 75005 Paris, FRANCE}
\vskip .2in

\end{center}
\date\today

\vskip 8pt


\vspace{0.5cm}
\begin{center}
{\bf Abstract}
\end{center}
\vspace{0.5cm}
The Fluctuation Relation  for a stationary state, kept at constant energy
by a {\em deterministic} thermostat - the Gallavotti-Cohen Theorem ---   
 relies on the ergodic properties of the system considered. 
We show  that when  perturbed by an energy-conserving  random noise,
the relation follows trivially for any system at  finite noise amplitude.
The  time needed  to achieve stationarity may
stay finite as the noise tends to zero, or it may diverge. In the former case the 
Gallavotti-Cohen result is recovered,  
while in the latter case, the crossover time may be computed
from the action of `instanton' orbits that bridge attractors and repellors. 
We suggest that the `Chaotic Hypothesis' of Gallavotti can thus be reformulated as a 
matter of  stochastic stability of the  measure in trajectory space. In this form this 
hypothesis may be directly tested. 

\vspace{1cm}


\end{titlepage}

\vspace{.5cm}
\section{Introduction}
\setcounter{equation}{0}
\renewcommand{\theequation}{\thesection.\arabic{equation}}
\label{Introduction}
\vspace{.5cm}

The Fluctuation Theorems  are relations satisfied by the distribution 
of entropy production ${\bar{\sigma}}_\tau=\frac{1}{\tau}\int_o^\tau dt \; \sigma(t)$~\cite{EvCoMo,EvSe,GaCo3,FT} 
\begin{equation}
\frac{ P({\bar{\sigma}}_\tau)}{P(-{\bar{\sigma}}_\tau)} \sim 
e^{ \tau {\bar{\sigma}}_\tau} \qquad 
{\mbox{or}}
\qquad 
\langle e^{-\lambda \tau {\bar{\sigma}}_\tau} \rangle = 
\langle e^{-(1-\lambda) \tau {\bar{\sigma}}_\tau} \rangle   \quad \forall \lambda
\label{ft}
\end{equation}
They hold in a series of different settings. The
 proofs are very simple and general when they apply to systems
starting from an equilibrium configuration~\cite{EvSe},
 or in contact with a thermal bath
involving stochastic noise that guarantees ergodicity~\cite{Ku,LeSp}.  
On the contrary, when the system is treated in the stationary state 
achieved thanks to a {\em deterministic} thermostat, the derivation is 
not simple and 
requires a  knowledge of the ergodic properties that is very rarely 
accessible analytically~\cite{GaCo3,comparison}.

In those cases in which the nature of the thermostat is thought to be physically
irrelevant, it is quite natural to
 prefer the stochastic baths 
because they invoke the least hypotheses, 
  making the analysis conceptually simpler.
On the other hand, the very fact that for deterministic systems
the Fluctuation Relation is not automatically satisfied makes it an interesting tool
to study questions of ergodic theory that are still largely open.   

For a purely deterministic  case,
 a forced system converges to an  `attractor' set, while the time-reversed 
dynamics converges to a `repellor' set.
Loosely speaking, the
 Gallavotti-Cohen theorem is then based on two types of hypotheses: {\em i)}
that the  attractor is sufficiently chaotic,  and {\em ii)}
 that the attractor and repellor are interwoven fractals 
with overlapping closures. 
As the forcing is increased,
 attractor and repellor tend to separate, and even if
the attractor 
remains chaotic the Fluctuation Relation no longer holds in general
\cite{BoGa}.
For macroscopic systems at reasonable levels of forcing, one can assume that
the hypotheses above are effectively (if not always strictly) satisfied, leading to 
 the Chaotic Hypothesis~\cite{ChH}.

In this paper we recast the content of the previous paragraph in the language
of stochastic stability~\cite{LSY}:
 we first show that adding a small, energy-conserving noise
the Fluctuation Relation is trivially satisfied by any model,
 and we then analise the limit
of  noise going to zero. Indeed, all 
the subtelties of the Gallavotti-Cohen theorem and
 of the Chaotic Hypothesis are a result of this limit.  The small noise limit
has moreover  the advantage that  an approximation scheme 
(analogous  to the semiclassical treatment) becomes exact, so that one can analise it
in detail. 

\section{Thermostatted dynamics}

We shall consider Hamilton's equations with forcing, a
 thermostatting mechanism, and energy conserving noise ~\cite{TaKu}:
\begin{eqnarray}
\dot q_i&=&p_i \nonumber \\
\dot p_i&=&-\frac{\partial V({\bf q})}{\partial q_i}+g_{ij} \eta_j 
- f_i({\bf q}) + \gamma(t) p_i = -\frac{\partial V({\bf q})}{\partial q_i}
+ g_{ij} (\eta_j - f_j)
\label{eq1}
\end{eqnarray}
here and in what follows we assume summation convention.
\begin{itemize}
\item $\eta_j$ are white, independent noises of variance $\epsilon$.
\item $g_{ij}=\delta_{ij}-\frac{p_i p_j}{{\bf p^2}}$ is the projector
onto the space tangential to the energy surface.
\item $f_i({\bf q})$ is a forcing term that depends on the coordinates only.
\item $\gamma(t) = \frac{{\bf f}\bullet {\bf p}}{{\bf p^2}}$.
 Multiplying the first of (\ref{eq1}) by 
$\frac{\partial V({\bf q})}{\partial q_i}$, the second
by $p_i$ and adding, one concludes that energy is conserved.

\item  The product 
$g_{ij}\eta_j$ is rather ill defined because both terms depend on time
and $\eta_j$ is discontinuous. The ambiguity is raised by
stipulating that it has to be interpreted in the Stratonovitch convention~\cite{Risken},
the meaning of which will be made clear below.
\end{itemize}
The probability evolves through~\cite{Risken}:
\begin{equation}
\dot P= -HP \label{eq2}
\end{equation}
where $H$ is the operator:
\begin{equation}
 H= p_i \frac{\partial}{\partial q_i}-\frac{\partial V({\bf q})}{\partial q_i}
\frac{\partial}{\partial p_i}+ \frac{\partial}{\partial p_i}\; [\gamma p_i]
- \frac{\partial}{\partial p_i} f_i - \epsilon \frac{\partial}{\partial p_j}
g_{ij} g_{il} \frac{\partial}{\partial p_l} 
\label{eq3}
\end{equation}
 The precise factor ordering in the last term is exactly what we meant by
Eq. (\ref{eq1}) being in the `Stratonovitch convention'.
 In the absence of driving ${\bf f}=0$ it is easy to check that $H$
anihilates any function that depends on the phase-space coordinates 
only through the energy $E=\frac{{\bf p}^2}{2}+V $. 
Hence, the noise respects the microcanonical
 measure, in that case.

We shall  consider as usual the `time-reversal transformation'
$\tilde H = \left[Q H Q^{-1} \right]^\dag$,
 where $Q$ is the operator that reverses 
momenta: $Q p_i Q^{-1} =  -p_i$ and 
$Q \frac{\partial}{\partial p_i} Q^{-1} = -\frac{\partial}{\partial p_i}$.
Under momentum reversal 
 all terms
change signs, except the last. Under Hermitean conjugation the factor order is reversed,
and derivatives change sign. We get:
\begin{eqnarray}
\tilde H =\left[ Q H Q^{-1}\right]^\dag &=& H + \sigma  \nonumber\\ 
               \sigma  &\equiv &  [\gamma p_i]\frac{\partial}{\partial p_i}\
-\frac{\partial}{\partial p_i}\; [\gamma p_i] =  -(d-1) \gamma 
\label{dodo}
\end{eqnarray}
with $2d$ is the dimension of phase-space,
which can be checked by  straightforward calculation.
We hence have:
\begin{eqnarray}
\left[ Q H Q^{-1}\right]^\dag &=& H + \sigma \\
\left[ Q (H+\lambda \sigma) Q^{-1}\right]^\dag &=& H + (1-\lambda)\sigma
\end{eqnarray}
because the entropy production rate $\sigma$ changes sign under 
time-reversal and conjugation.


From here it is standard to show that for every finite value of
$\epsilon$ the fluctuation theorem holds \cite{Ku,LeSp}.  For completeness, let us
 outline 
the proof:
given any initial and final configurations $x_i=(q_i,p_i)$ and  $x_f=(q_f,p_f)$
the  expectation of the total work is
\begin{eqnarray}
\langle e^{-\lambda \int_o^\tau \sigma dt} \rangle &=&
 \langle  x_f |  e^{-\tau(H + \lambda \sigma)} |x_i \rangle \nonumber \\
&=&   \langle  x_i |  e^{-\tau(H + \lambda \sigma)^\dag} |x_f \rangle
= \langle  x_i |  Q^{-1} e^{-\tau(H + (1-\lambda)  \sigma)} Q |x_f \rangle
\label{coco}
\end{eqnarray}
Because of the noise, the evolution is ergodic on the energy surface, and thus
the real part of the spectrum of $H$ restricted to functions on the
 energy surface has a gap.
Evaluating (\ref{coco}) on  a basis of eigenfunctions on the energy surface,
the `lowest' right and left eigenfunctions (the ones 
with the eigenvector $\mu_o(\lambda)$ having the lowest real part)  $\langle L_o|$ and 
$| R_o \rangle $  dominate:
\begin{eqnarray}
 \langle  x_f |  e^{-T(H + \lambda \tau \sigma)} |x_i \rangle &\sim&
   \langle  x_f | R_o \rangle \langle L_o|x_i \rangle \; e^{-\tau \mu_o(\lambda)}
\nonumber \\
&=&   \langle  x_f | L_o \rangle \langle R_o|x_i \rangle \; e^{-\tau \mu_o(1-\lambda)}
\end{eqnarray}
and using the fact that $H + \lambda \sigma$ and $H + (1-\lambda)  \sigma$ have the same
spectrum, we conclude that:
\begin{equation}
\lim_{\tau \rightarrow \infty}
 \left\{ \frac{1}{\tau} \ln  \langle e^{-\lambda \int_o^T \sigma dt} \rangle -
\frac{1}{\tau} \ln  \langle e^{-(1-\lambda) \int_o^T \sigma dt} \rangle \right\}= 0 
\end{equation}
which is the Laplace transformed version of the Fluctuation Theorem (\ref{ft}).
Clearly, the whole argument breaks down at strictly zero noise, when 
the eigenfunctions on the energy surface are no longer smooth, and the 
overlaps $\langle L_o|x_f  \rangle $ and  $\langle R_o|x_i \rangle $
may vanish.
 The problems thus may (and will) arise
because we have proven the theorem for
\begin{equation}
\lim_{\epsilon \to 0} \lim_{\tau \to \infty} \;\; \frac{1}{\tau} \;
 \ln P\left( \int_o^\tau dt \;\sigma \right)
\end{equation}
and the Gallavotti-Cohen theorem is for:
\begin{equation}
\lim_{\tau \to \infty} \lim_{\epsilon \to 0}  \;\; \frac{1}{\tau} \; 
\ln P\left( \int_o^\tau dt \;\sigma \right)
\end{equation}
In what follows, we shall
 analise the small-noise limit, starting from a simple example.

\section{A simple example}

 Consider a particle on a two-dimensional space with
 toroidal boundary conditions, a constant field E along the direction $1$,
 and an isokinetic thermostat. The example was 
discussed in Ref. \cite{comparison}. 
We write the velocity vector $(p_x,p_y)$ as:
\begin{equation}
p_x=p \cos \theta \;\;\; ; \;\;\; p_y=p \sin \theta
\end{equation}
There are two stationary situations:  parallel velocity $(p_x,p_y)=(p,0)$, $\theta=0$
(stable: the attractor) and antiparallel  velocity $(p_x,p_y)=(-p,0)$, $\theta = \pi$
(unstable: the repellor).
Equation (\ref{eq1} ) reads, in this case:
\begin{eqnarray}
\dot p_x &=& E - \frac{E p_x^2}{p^2} + \eta_x - p_x 
\frac{p_x \eta_x + p_y \eta_y}{p^2} 
\nonumber \\
\dot p_y &=&  - \frac{E p_x p_y}{p^2}  +  \eta_y - p_y
\frac{p_x \eta_x + p_y \eta_y}{p^2} 
\end{eqnarray}
In terms of $\theta$
both equations collapse into a single one:
\begin{equation}
\dot \theta = -E \sin \theta + \eta_\theta = -\frac{d}{d\theta} [-E \cos \theta]
+ \eta_\theta
\label{vu}
\end{equation}
where we have defined the angular, isotropic  noise: 
$ \eta_\theta \equiv \sin \theta \eta_x - \cos \theta \eta_y$, a white noise
 whose variance is $\epsilon$. 
The entropy production rate is $\sigma = E \cos \theta$.

\subsection{Weak noise limit}

\begin{figure}[ht]
\vspace{-2cm}
\begin{center}
  \includegraphics[width=.5\columnwidth]{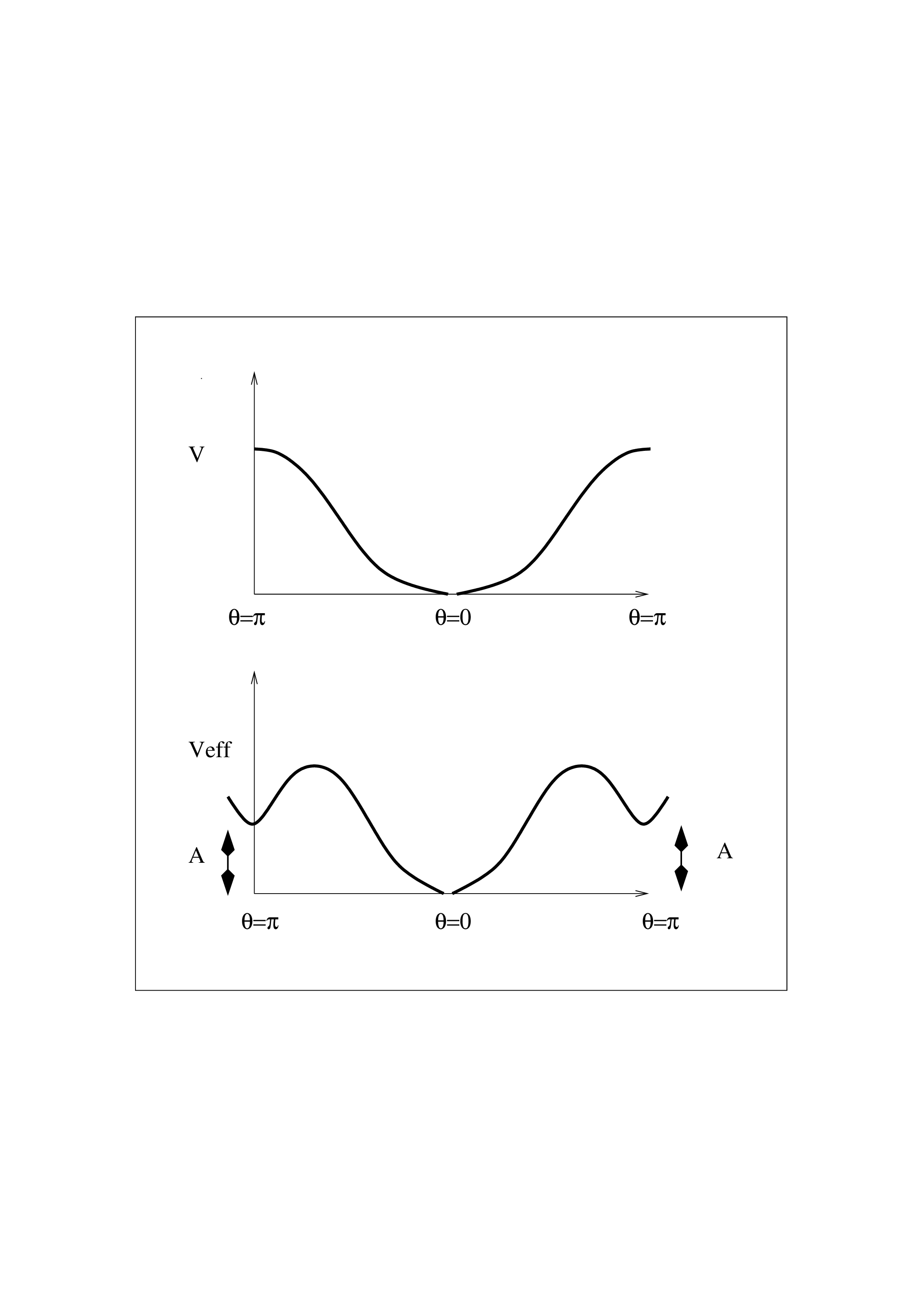}
\end{center}
\vspace{-2cm}
\caption{Top: potential in terms of angle. Bottom: effective potential 
associated with the Shroedinger problem (\ref{pp}). $A$ is the subdominant
bias proportional to $(\lambda-1/2)$ that favours attractor $(A>0)$,
or repellor $(A<0)$. }
  \label{fig2}
\end{figure}

Equation (\ref{vu}) can be interpreted as a Langevin process in an effective
 potential $V=-E \cos \theta$, see Fig.  \ref{fig2}. 
The probability evolves according to:
\begin{equation}
\dot P = - \left[ \epsilon \frac{\partial^2}{\partial \theta^2}
+ E \frac{\partial}{\partial \theta}  \sin \theta \right] \; P
= - H_{FP} P
\end{equation}
If we are interested in calculating the average  $e^{-\lambda \int \sigma dt}$
in the large-time limit we need to compute the lowest eigenvalue
of 
\begin{equation}
H_\lambda = H_{FP} + \lambda E \cos \theta 
\end{equation}
which can be taken as usual to the Hermitian form 
\begin{equation}
H_\lambda^h= e^{\epsilon E \sin \theta} H_\lambda 
e^{-\epsilon E \sin \theta} = \frac{1}{\epsilon} \left[
-  \epsilon^2 \frac{\partial^2}{\partial \theta^2} + \frac{E}{4} \sin^{2} \theta +
\epsilon E 
(\lambda-\frac{1}{2}) \cos \theta \right]
\label{pp}
\end{equation}
and takes the form of a
 Shroedinger operator in an effective potential 
$V_{eff}=  \frac{E}{4} \sin^{2} \theta 
+\epsilon  (\lambda-\frac{1}{2}) E \cos \theta $
depicted in the figure \ref{fig2}.
The average work per unit  time at long times is given by
the lowest eigenvalue of (\ref{pp}). The Gallavotti-Cohen symmetry
 is manifest making $(\lambda-1/2) \rightarrow
-(\lambda-1/2)$ and $\theta \rightarrow \pi - \theta$.

Now, in the limit of small $\epsilon$ the situation is as follows: to leading
order in $\epsilon$ the ground state is doubly degenerate, the eigenfunctions being
concentrated in $\theta=0$ and $\theta=\pi$, respectively --
 both attractor and repellor are selected by the potential $V_{eff}$ at this order
 ($\epsilon^{-1}$). 
It is the next-to-leading order   $A=(2\lambda-1)E \cos\theta$ in $V_{eff}$ 
(see Fig. \ref{fig2}) that lifts the degeneracy by a quantity of $O(1)$ in $\epsilon$,
 thus selecting
attractor or repellor, depending on the sign of $ (\lambda-1/2)$.
Clearly,  if  $(\lambda-1/2)$ is of order one, for small $\epsilon$  
the ground state eigenfunction is peaked in  only one well, which one depends
on the sign of  $(\lambda-1/2)$, and :
\begin{equation}
\langle e^{-\lambda \tau {\bar{\sigma}}_\tau}\rangle  \sim 
e^{\tau (| \lambda-1/2| -1/2) E}
\label{uuu}
\end{equation}
where the last term $1/2$ comes from the zero-point 
energy of the minimum.

\subsection{Attractor, repellor and time-reversal}

\begin{figure}[ht]
\vspace{-2cm}
\centering
  \includegraphics[width=8cm]{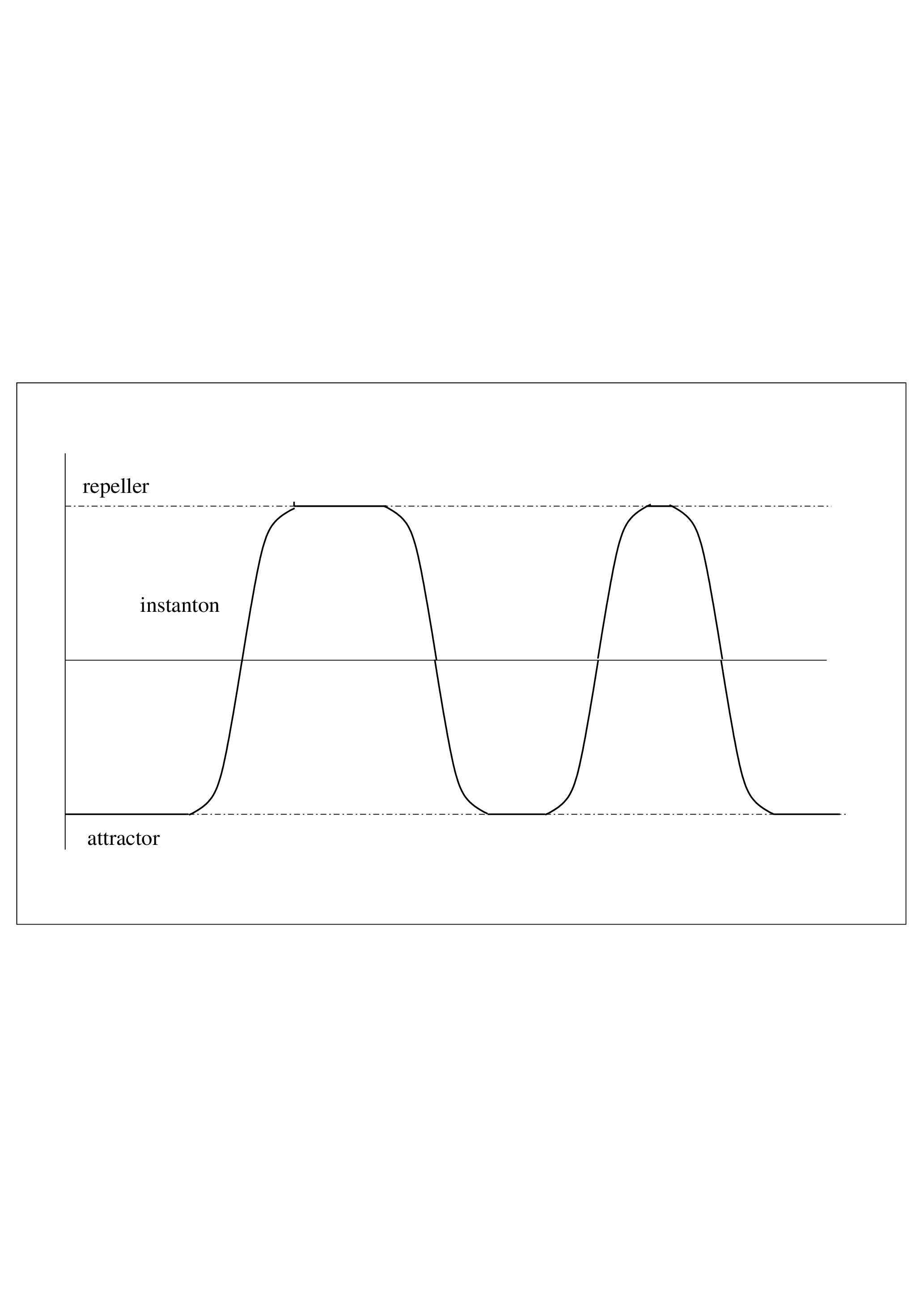}
 
 \vspace{-2cm}
 
 \caption{An `instanton' - `antiinstanton' gas configuration. The average time
spent in the attractor and repellor determines the entropy production.}
  \label{inst}
\end{figure}

Let us try now a more direct approach (the techiques can be found in \cite{ZJ}).
We can write an expression for the probability of a trajectory
in configuration space:
\begin{equation}
P({{traj}})= e^{-\frac{1}{\epsilon}\int \; dt\; \left[
 \frac{1}{4}(\dot \theta + E \sin \theta)^2 - \frac{1}{2} E \epsilon \cos \theta\right]}
\label{un}
\end{equation}
and we wish to calculate the average
\begin{equation}
\langle e^{-\lambda \sigma} \rangle =
\int_{{traj}} \; P({{traj}}) \; e^{-\lambda \int dt \;\cos \theta}
=\int_{{traj}} 
 e^{-\frac{1}{\epsilon}\int \; dt\; \left[
 \frac{1}{4} (\dot \theta + E \sin \theta)^2 + (\lambda -\frac{1}{2}) E \epsilon 
 \cos \theta\right]}
\label{do}
\end{equation}
The trajectory weight in (\ref{do}) is composed of two terms, the leading term
corresponding to the Lagrangian:
\begin{equation}
L= \frac{1}{4} (\dot \theta + E \sin \theta)^2 = \frac{1}{4} \dot \theta^2 + 
\frac{1}{4}  E^2 \sin^2 \theta - \frac{1}{2} E 
\frac{d \cos \theta}{dt}
\label{L}
\end{equation}
and a subleading term proportional to $(\lambda -1/2)$.
The dominant terms must satisfy Lagrange's equations derived from (\ref{L}): these
correspond to a problem with inertia in the {\em inverted potential} $-V_{eff}$.  
The solutions  are:
\begin{equation}
 \dot \theta = \pm \sqrt{4 {\cal{E}} +  E^2 \sin^2 \theta} 
\label{EL}
\end{equation}
where $ {\cal{E}}$ is a constant playing the role of an energy associated with the
 Lagrangian (\ref{L}).
It is easy to see that the integral in the exponent (\ref{un}) (the action) 
can only be 
finite for large times if the trajectory spends most of the time in attractor 
or repellor, and this is only possible if ${\cal{E}}=0$. We thus have that solutions
are either solutions of the original noiseless problem:
\begin{equation}
 \dot \theta = 
-E \sin \theta = -\frac{d}{d\theta} [-E \cos \theta]
\end{equation}
which converge to the attractor and have zero action, or `rare':
\begin{equation}
 \dot \theta = 
E \sin \theta = \frac{d}{d\theta} [-E \cos \theta]
\end{equation}
taking from attractor to repellor, and these have a finite action equal to 
$E\int d \theta \; \dot \theta \sin \theta= 2E$.
The general trajectory selected by the first term of (\ref{do}) is composed of
sejours in the attractor and in the repellor, with rapid descents from repellor
to attractor  (all three with zero action), plus the `instanton' 
trajectories with action $2E$, see Fig. \ref{inst}.

An important point to note is that the trajectories associated with 
the Lagrangian (\ref{L}) in the noiseless limit 
are, as we have mentioned, many more than the  trajectories of the original
dynamic system in the absence of noise - they include all the non-zero action solutions.
Also very important is the fact that  the dynamics associated with (\ref{L})
 differs from the original dynamics in that the original stable point $\theta=0$ 
is now  unstable, due the  reversal of the potential.
These questions will be relevant in the following sections.

\begin{figure}[ht]
\vspace{-2cm}
\centering
  \includegraphics[width=8cm]{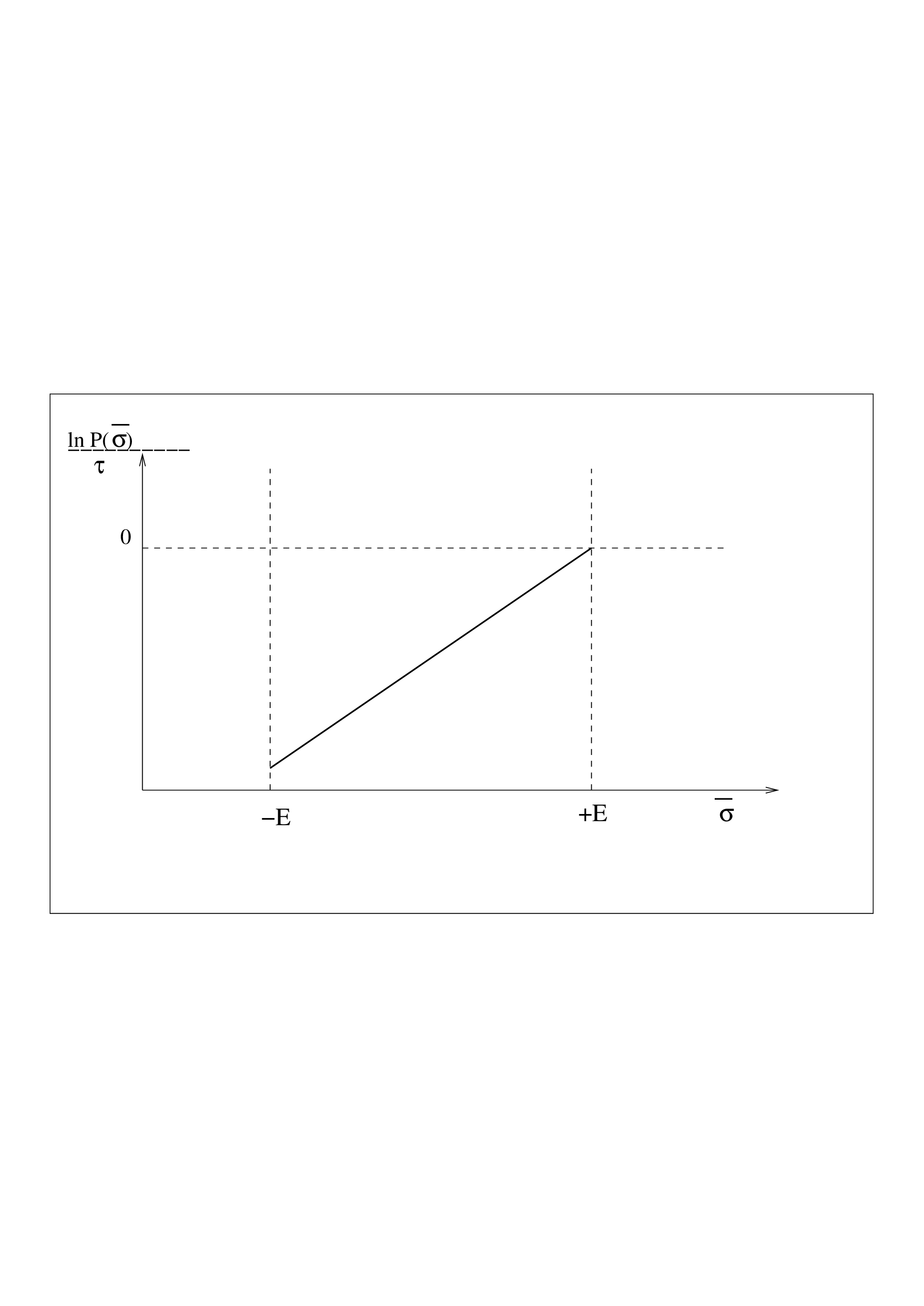}

\vspace{-2cm}

\caption{$\ln P(\sigma)$ vs. $\sigma$ for the problem of section II.
The profile is linear, and can be seen as the coexistence curve of
two phases with free-energy $\pm E$. }
  \label{li}
\end{figure}

The exact proportion of time spent in attractor $\tau_{{attractor}}$
and repellor $\tau_{{repellor}}$ will be determined by
the second, subdominant term.
This term then gives an action  $\sim \epsilon E (\lambda -1/2)
 (\tau_{{repellor}} - \tau_{{attractor}})$.
The dominant contribution for the $\lambda$-dependent average 
is given for large times $\tau$ by the trajectories  that maximise
\begin{equation}
- \ln P({{traj}}) \sim  - \epsilon (\lambda -1/2)
(  \tau_{{repellor}} - \tau_{{attractor}}) +
N_{{instanton}} \Delta/\epsilon
\label{action} 
\end{equation}
where $N_{{instanton}}$ 
is the number of climbings from attractor to repellor.
If $\lambda=0$ the system is unbiased and it
  will prefer to stay in the attractor. If, on the contrary, we
wish to compute the large-deviation function with $\lambda \neq 0$, then
the trajectories that dominate will share time in attractor and repellor,
so that:
\begin{equation}
 (\tau_{{repellor}} +  \tau_{{attractor}})
{\bar{\sigma}}_\tau= \tau_{{repellor}} {\bar{\sigma}}_{{repellor}} +  \tau_{{attractor}}
 {\bar{\sigma}}_{{attractor}} =( \tau_{{attractor}}- \tau_{{repellor}}) E
\end{equation}
For the probability of work  per unit time  $\bar \sigma$ , at small noise level,
 we get a linear profile as in figure \ref{li}

which is in agreement with (\ref{uuu}).

 For infinite times the last term
in (\ref{action}) is negligible,
 but for times such that instanton configurations weigh comparably
to the terms proportional to time: $\Delta/\epsilon \sim \ln \tau$
and we shall not see reversals.
There are then two possibilities:  we may consider a set of initial 
conditions corresponding to an  equilibrium configuration, and get 
the correct proportion of sejour in attractor and repellor {\em without 
 needing to perform any activated jumps from attractor to repellor}.
This is the transient fluctuation theorem, which  holds for
zero noise. On the other hand, if we consider any other initial configuration,
we need jumps from attractors to repellor and back in order to obtain the 
right proportion of sejours in the large time limit. Clearly, this second 
`stationary' fluctuation theorem needs  $\ln \tau \gg \Delta/\epsilon  $.

Let us note  the close analogy  to a one-dimensional
Ising model~\cite{BoGa} $E=  - J \sum_i s_i s_{i+1} - h \sum s_i$  with
$\beta J \rightarrow \exp(\Delta/\epsilon)$  the wall energy
(where $\Delta$ is the instanton
action), and $\beta h \rightarrow   \epsilon (\lambda - 1/2)$.
Although in our case `climbing' walls cost, and `descending' ones do not,
the difference is unimportant because the number of instantons and of
antiinstantons differ by at most one.
Here $\epsilon$ enters in  the length needed to
accomodate many walls, which is necessary in the thermodynamic 
(large-length) limit. The alternative is to consider a properly weighted
set of boundary conditions, so that no walls are necessary: this is
analogous to the transient fluctuation theorem.

\section{Small noise limit, general}

We shall now show that the structure of the previous section is 
quite general, although  there are many instanton-antiinstanton
trajectories.
Consider the evolution equations (\ref{eq2}) and (\ref{eq3}).
We write this as a path integral
\begin{equation} 
P({\bf q},{\bf p}) =\int_{[{{paths}}]} 
e^{-\frac{\Delta_{{path}} }{\epsilon}
}= \int_{[{{paths}}]} e^{-\frac{1 }{\epsilon}
\int dt \; \left( L({\bf q},{\bf p})- \frac{\epsilon}{2} \sigma \right) }
\end{equation}
where the paths have the appropriate initial condition in phase-space,
 and all possible final configurations.
\begin{eqnarray}
L&=&   \hat q_i (\dot q_i-p_i) + \hat p_i (\dot p_i +  
V_i + g_{ij}f_j) + 
g_{ij} \hat p_j g_{il} \hat p_l \nonumber \\
&=&  \hat q_i \dot q_i + \hat p_i \dot p_i -  H
\end{eqnarray}
with
\begin{equation}
 H=    \hat q_i p_i - \hat p_i  g_{ij} f_j - 
\hat p_i   V_i-
g_{ij} \hat p_j g_{il} \hat p_l 
\end{equation}
Here and in what follows we use the notation 
$V_i=\frac{\partial V}{\partial q_i}$.
The second, subdominant 
 term in the exponent comes from the symmetrisation (cfr. the second
of (\ref{dodo})):
\begin{equation}
\frac{\partial}{\partial p_i}(\gamma p_i)=\frac{1}{2}\left[
\frac{\partial}{\partial p_i}(\gamma p_i)+
(\gamma p_i)\frac{\partial}{\partial p_i}\right] - \frac{\sigma}{2}
\end{equation}
where again $\sigma=-(d-1)\gamma$.

We wish to calculate the large-deviation function
\begin{equation}
\langle e^{\lambda \int_o^\tau dt \; \sigma }\rangle = \int_{[{{paths}}]} 
e^{-\frac{1}{\epsilon}
\int dt \; [L({\bf q},{\bf p}) + \epsilon (\lambda-\frac{1}{2})  \sigma ] }
\label{pth}
\end{equation}
Just as in the previous section, in the limit of small noise we have to
look for saddle-point solutions only taking into account the first
term in the exponent.
The  equations of motion then read:
\begin{eqnarray}
\dot{\hat q}_i &=& -\frac{\partial H}{\partial q_i}= \hat p_l  g_{lr}
 \frac{\partial f_r}{\partial q_i} +
 \hat p_l  \frac{\partial V_l}{\partial q_i } 
\nonumber \\
\dot{\hat p}_i &=&  -\frac{\partial H}{\partial p_i}= -\hat q_i - 
\left[ \hat p_l f_j + 
+ \hat p_l \hat p_j\right]  \frac{\partial g_{lj}}{\partial p_i}
 \nonumber \\
\dot{ q}_i &=& \frac{\partial H}{\partial {\hat q}_i}= p_i\nonumber \\
\dot{ p}_i &=&  \frac{\partial H}{\partial {\hat p}_i}=
 -g_{ij} (f_j +2\hat p_j)-V_i
\label{eqs}
\end{eqnarray}

Eqs. (\ref{eqs}) are Hamilton's equations, and they involve twice as many
degrees of freedom as the original system. In fact, the original
phase space variables $(q_i,p_i)$ have now become
the coordinates, and their conjugate momenta are the new variables  
$(\hat q_i,\hat p_i)$.Multiplying the
 third equation by $V_i$, the fourth by $p_i$ and adding, we conclude
that {\em all} the solutions in the extended phase-space of $(\ref{eqs})$ 
are still thermostatted to $\sum p_i^2+V=constant$.

Let us now rewrite the action in Lagrangian form.
We first put $\hat p_i  g_{ij} f_j = \hat p_l g_{li}  g_{ij} f_j$
where we have used $g_{li}g_{ij}=g_{lj}$. Next, we separate tangential
and normal parts of $\hat p_i \dot p_i = \hat p_l g_{li} \dot p_i +
\hat p_l p_l \; \dot p_i p_i/p^2$.
We can complete squares and collect the terms, to get the Lagrangian:
\begin{eqnarray}
 &L& = \hat q_i(\dot q_i - p_i) -\frac {1}{4}
   \left[\dot p_i + g_{ij}f_j+ V_i \right]^2     + \nonumber \\
& & \left[ 
\frac{1}{2} 
\left(\dot p_i + g_{ij}f_j+  V_i
\right)+ 
g_{ij} \hat p_j 
\right]^2 +
 \frac{\hat p_l p_l}{p^2} \;\left( \dot p_i p_i + p_i V_i \right)
\end{eqnarray}
Using the last two equations of motion (\ref{eqs}), we see that at the saddle 
point level the last two terms cancel. We hence
conclude that the action to leading order equals 
\begin{equation}
\Delta= \int dt \; \tilde L(\ddot{\bf q},\dot{\bf q},{\bf q})= \int dt \;
\left[\ddot q_i + g_{ij} f_j+ V_j  \right]^2
\label{quadr}
\end{equation}
where we have substituted $p_i=\dot q_i$ everywhere in $\tilde L$. 
Note that $\tilde L$ has an  unusual second time derivative $\ddot q_i$.
All relevant trajectories are the saddle points of the action (\ref{quadr}):
\begin{equation}
\frac{d^2}{dt^2} \frac{\partial \tilde L}{\partial \ddot q_i} 
-\frac{d}{dt} \frac{\partial \tilde L}{\partial \dot q_i}
+ \frac{\partial \tilde L}{\partial  q_i}=0
\label{ee}
\end{equation}
The solutions are determined by initial and final values of {\em both}
$q_i$ and $\dot q_i$.
In particular, 
 trajectories of the original system without noise are the ones  
having zero action:
\begin{equation}
\ddot q_i + g_{ij}f_j +  V_i=0
\label{eqqq1}
\end{equation}
Furthermore, in order that the action be finite, we must have that 
$\ddot q_i + g_{ij}f_j  + V_i\rightarrow 0$ for large $t$.
As an example, in the next section we shall construct the 
solutions of Eq. (\ref{ee})
for the Lorentz gas.

In a completely chaotic system, we can  construct an expansion
in terms of (isolated) trajectories~\cite{book,Creagh}, and just as before 
these will commute between attractor(s) and repellor(s) -- defined as
dynamically stable and unstable solutions 
of the equations of motion of the original system -- where they will
spend essentially all the time, since otherwise their action would be infinite.
Because the only difference between attractors and repellors is, by definition,
their stability, the next-to-leading contribution in the action
\begin{equation}
\Delta_\epsilon = \left(\lambda-\frac{1}{2}\right) \int dt \; \sigma + \frac{1}{2}
\int dt \; dt' \; 
\left. \frac{\delta^2 \tilde L}{\delta q_i(t) \delta q_j(t')}\right|_{traj} \;
 \delta q_i(t) \delta q_j(t')
\label{ntl}
\end{equation}
 will
decide the length of the sejours in each for each $\lambda$.

In this setting, a necessary condition for the Gallavotti-Cohen theorem 
(i.e. the Fluctuation relation at zero noise) to hold, is that there be orbits
 with zero action bridging attractor and repellor. 

\subsection{Variational solutions: the zero-action  situation.}

Given a point on the attractor $(q_a,p_a=\dot q_a)$ and a point on the repellor
 $(q_r,p_r=\dot q_r)$ and two times $t_i$ and $t_f$ we can obtain the solution
going from   $(q_a,\dot q_a)$ at  $t_i$ to  $(q_r,\dot q_r)$ at
$t_f$ as follows.
Starting from any curve $({\bf q}(t), {\bf \dot q}(t))$,
contained within the energy surface and
having the correct endpoints, its action will be an upper bound
for the true, minimal action. Deforming the curve
 so as to minimize the action
 (\ref{quadr}) we reach a solution.
 The arrival time may become a 
variational parameter itself.

If the lowest  action $\Delta_{{min}}$
 associated with any trajectory joining any two points in the   
attractor and repellor is finite, we know that the system will
necessarily  need a time at least $\tau \sim 
 e^{\Delta_{{min}}/\epsilon}$ to 
exlore them. This can be simply seen by looking at the sum over 
paths (\ref{pth}) as a partition function in a space of trajectories
at inverse temperature $\epsilon$,
with  $\Delta_{{min}}$ its `ground state energy'.

Clearly, for the Gallavotti-Cohen theorem to apply, a necessary condition
 is that $\Delta_{{min}}=0$. For this to happen,
 attractor and repellor have to be  interwoven so that
 just advancing from  $(q_a,\dot q_a)$ along a zero-action  trajectory
satisfying (\ref{eqqq1}) we reach a point that is as close as we wish 
 to  some $(q_r,\dot q_r)$, providing  a zero-action `bridge'.
However, the condition  $\Delta_{{min}}=0$ only implies that
the time for bridging attractor and repellor  grows in a  subexponential ---
 for example power law $\tau \sim \epsilon^{-\mu}$ --- way, and is not necessarily
finite as $\epsilon \rightarrow 0$.
One can imagine such a situation arising in a case in which attractor and 
repellor have closures of dimension $d_1$, but these
intersect in a manifold of dimension $d_2<d_1$.
Without noise a trajectory might have vanishing probability of leaving 
the attractor and entering the repellor,
 since the time spent in the overlapping region would be negligible;
the addition of noise will coarsen both attractor, repellor and intersection
by introducing a width, hence making the passage more likely.

\section{Instantons for the Lorentz gas}

We shall consider here a Lorentz gas~\cite{Lorentz}, a system 
 with a single thermostatted 
particle under the action of
an electric field  as in section II, but with  in addition  obstacles
where the  particle bounces, introducing  chaos
The Lagrangian (\ref{L})
reads
\begin{eqnarray}
\tilde L&=&  \frac{1}{4}(\dot \theta + E \sin \theta  )^2 =
 p_\theta \dot \theta -
{\cal{H}}
\nonumber \\
{\cal{H}} &=&  {\cal{E}} =
2p_\theta \left( p_\theta - E \sin \theta \right)
\end{eqnarray} 
with $ {\cal{E}}$ defined in (\ref{EL}) and
\begin{equation}
p_\theta =  \frac{1}{2} (\dot \theta + E \sin \theta )
\end{equation}
The motion is punctuated by bounces on the wall.
Between bounces, the trajectories satisfy (\ref{EL}):
\begin{eqnarray}
 d \theta &=& \pm \sqrt{ 4{\cal{E}} +  E^2 \sin^2 \theta} dt \nonumber \\
  dx&=& \cos \theta dt = \frac{\cos \theta}{\dot \theta} d\theta =
 \pm \frac{\cos \theta}{\sqrt{ E^2 \sin^2 \theta +4 {\cal{E}}}}d\theta
\nonumber \\
dy&=& \sin \theta dt = \frac{\sin \theta}{\dot \theta} d\theta =
\pm \frac{\sin \theta}{\sqrt{ E^2 \sin^2 \theta + 4{\cal{E}}}}
d\theta
\end{eqnarray}
which yields the parametric equations for the trajectory segment.
One also has that
\begin{eqnarray}
\dot p_\theta &=& -2E  \;\cos \theta \; p_\theta = 2E  \; p_\theta  \;
 \dot x \nonumber \\
 p_\theta(t)  &=&  p_\theta(t_o)  \; e^{2E\; x|^t_o}
\end{eqnarray}
where $t_o$ is the initial time of the segment.
The action for the trajectory segment reads:
\begin{equation}
\Delta =  \int dt \; p_\theta^2(t) = p_\theta^2(t_o) \int dt \;e^{4E \; x|^t_o}
\label{ppp}
\end{equation}
where $x|^t_o=\int dt \; \dot x$ is the total distance along $x$, without
subtracting the windings if the space is periodic. It is positive if the particle moved
downhill, and negative if it moves contrary to the force.

Next, we need to consider bounces. Because  we are solving 
 the instanton dynamics, rather than the original one, we could wonder
what happens during a bounce. To make this point clear, in the Appendix we compute
a bounce by considering the wall as a limit of continuous  potentials.
The result is that:
{\em i)} the reflection law holds,  {\em ii)} the contribution of the bounce to the 
action vanishes in the hard wall limit,  and {\em iii)} $p_\theta$ does not 
change from just before to just after the bounce.
Instead, ${\cal{E}}$ is not conserved in bounces, since:
\begin{equation}
 {\cal {E}}|_{before}^{after} = \left. 2 p_\theta 
E \; \sin \theta\right|_{before}^{after}
\label{lll}
\end{equation}
which is nonzero in general, except when $p_\theta=0$.

Segments of trajectories belonging to the attractor and repellor of the original
noiseless dynamics
have   zero action and $p_\theta(t)= {\cal{E}}=0$:
 this is preserved by both trejectories
 and bounces at all times (cfr. Eqs. (\ref{ppp}) and
(\ref{lll})).

Trajectories have a starting point $(x,y,\theta,\dot \theta)$. 
Thanks to bounces,
they move uphill and downhill with respect to the field.
 Those that diffuse uphill have a smaller and smaller 
value of $p_\theta$ ( Eq. (\ref{ppp})). Furthermore, 
as  $p_\theta \rightarrow 0$ also ${\cal{E}} \rightarrow 0$ (although 
${\cal{E}}$ can still increase somewhat in a bounce).
Hence,  a trajectory that had bounces allowing it to diffuse a long way uphill
  becomes more and more closely
a  zero-action trajectory. Moreover,
 it is (or rather, it shadows) a trajectory belonging to the {\em repellor}, 
since trajectories in the attractor move on average downhill.

 Suppose next that the  starting point
 is a very small perturbation with $p_\theta \neq 0$ of a point on the
 attractor. At the beginning the point will continue to
 move on average downhill. However, this implies that the value of $p_\theta$
grows exponentially  ( Eq. (\ref{ppp})). 
In other words, in the full dynamics with instantons,
true attractor  trajectories are {\em unstable}, a general fact already
mentioned in the previous sections.
 After some time, these trajectories
will bounce  uphill and downhill. A few of  these perturbations
 will diffuse a large amount uphill, and as described above 
those  will then be very close to the repellor.
Hence, we have identified  the instantons interpolating between
attractor and repellor as the (rare) trajectories that have initial conditions
close to the attractor and 
such that they keep diffusing on average upwards. Their action is
just the integral of the exponential of the uphill distance - a finite 
quantity.

As is typical with instantons, we need some sort of
 `shooting' method to find those solutions that actually end up in (or shadowing an
orbit in) the repellor,
otherwise a generic solution will just evolve away.
Secondly, we see that there are many of these solutions, and not essentially one
as in the case without obstacles.

\pagebreak

\section{Conclusions}

We have recasted the problem of the deterministic Fluctuation Theorem as
the vanishing noise limit of stochastic one. Large-deviation functions in this limit
 need not coincide with the zero-noise result, but one can always postulate that this
is so in a practical  case, thus making an
{\bf alternative Chaotic Hypothesis~\cite{ChH}}:
Considering the energy-conserving noise described above (and more
generally noise respecting all the constants of motion of the  
dynamics),
{\em a reversible many-particle system in a stationary state 
can be regarded as stochastically stable 
for the purposes of computing probabilities over trajectories of 
macroscopic observables.}
The advantage of this formulation is that it is immediately testable
 in  a numerical (and perhaps also in a real) experiment.

We have argued here that if there are orbits with zero action that bridge 
attractor and repellor, then the time needed in order that the Fluctuation Relation
 is satisfied is
subexponential in the noise $\epsilon \ln \tau \rightarrow 0$. We have, however,
fallen short of proving the Gallavotti-Cohen theorem for that case, since
we argue that  subexponential yet slowly divergent growth of $\tau$ as 
$\epsilon \to 0$ 
  is likely to arise, perhaps in
 systems in which the closures of attractor and repellor are simultaneously dense
in a manifold of low dimension. It seems an interesting question
to obtain estimates of passage times in terms of the noise (or temperature)
 in such cases.   

\vspace{2cm}

{\bf Acknowledgemnts}

I wish to thank F. Zamponi for discussions and suggestions.

\pagebreak

\section{Appendix. Instantons and hard walls}

To model the bounce, we consider walls as regions with a constant repulsive
force (perpendicular to the wall), and then take the limit of large force.
We shall rotate the axis to a new axes $(x',y')$ so that
outside the wall region the field is $(E_1,E_2)$ and inside the wall
region the field is $(E^w,0)$, $E^w \rightarrow \infty$ as shown in the figure.

\begin{figure}[ht]
\vspace{-2cm}

\centering
  \includegraphics[width=8cm]{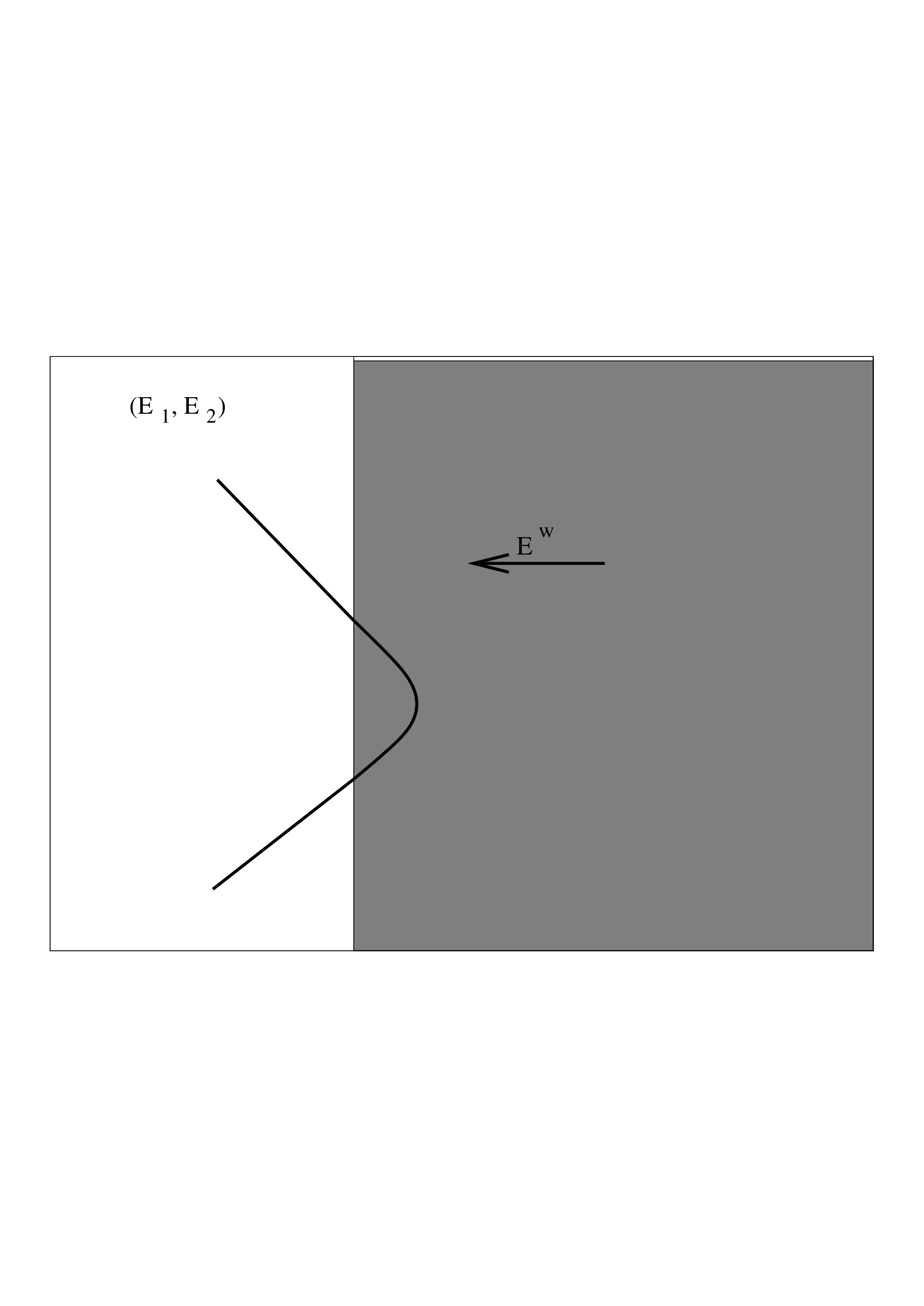}

\vspace{-2cm}

\caption{Bounce trajectory in a wall region.}
  \label{fig:Bounce in a wall}
\end{figure}

In general, in a region of fields $( E_1, E_2)$ 
the Lagrangian reads
\begin{eqnarray}
L&=&  \frac{1}{4}(\dot \theta + E_1 \sin \theta -  E_2 \cos \theta )^2 
\nonumber \\
{\cal{H}} &=& 2 p_\theta \left(p_\theta - E_1 \sin \theta
 +  E_2 \cos \theta \right)
\end{eqnarray}
with 
\begin{equation}
p_\theta =   \frac{1}{2} (\dot \theta + E_1 \sin \theta -  E_2 \cos \theta) 
\end{equation}
The forces $ E_i$ 
jump when the particle crosses the wall. Across this jump,
$\theta$ is continuous, but $\dot \theta$ jumps. To calculate this,
we write the equations of motion:
\begin{equation}
2 \frac{d}{ dt} (\dot \theta +E_1 \sin \theta - E_2 \cos \theta )
- 2(\dot \theta +E_1 \sin \theta - E_2 \cos \theta )
(E_1 \cos \theta + E_2 \sin \theta )=0
\label{oo}
\end{equation}
which implies that in regions of constant fields:
\begin{eqnarray}
 \frac{d}{ dt} \ln (\dot \theta +E_1 \sin \theta - E_2 \cos \theta )
&=&
2 (E_1 \dot x +  E_2 \dot y ) \nonumber \\
\ln \left|\dot \theta +E_1 \sin \theta - E_2 \cos \theta \right|^t_o
&=& \left. 2 (E_1 x +E_2 y)\right|^t_o = 
\left.  2 {\bf {x\bullet E}} \right|^t_o
\label{sol}
\end{eqnarray} 
On entering the wall region, equation (\ref{oo}) implies that  
\begin{equation}
p_\theta(t)=  \frac{1}{2}(\dot \theta +E_1 \sin \theta - E_2 \cos \theta )
\end{equation}
is continuous across discontinuities of the potential.
Instead, the numerical value of  $H={\cal{E}}$ jumps when the trajectory bounces,
except if $p_\theta=0$.

\begin{figure}[ht]
\vspace{-2cm}
\centering
  \includegraphics[width=8cm]{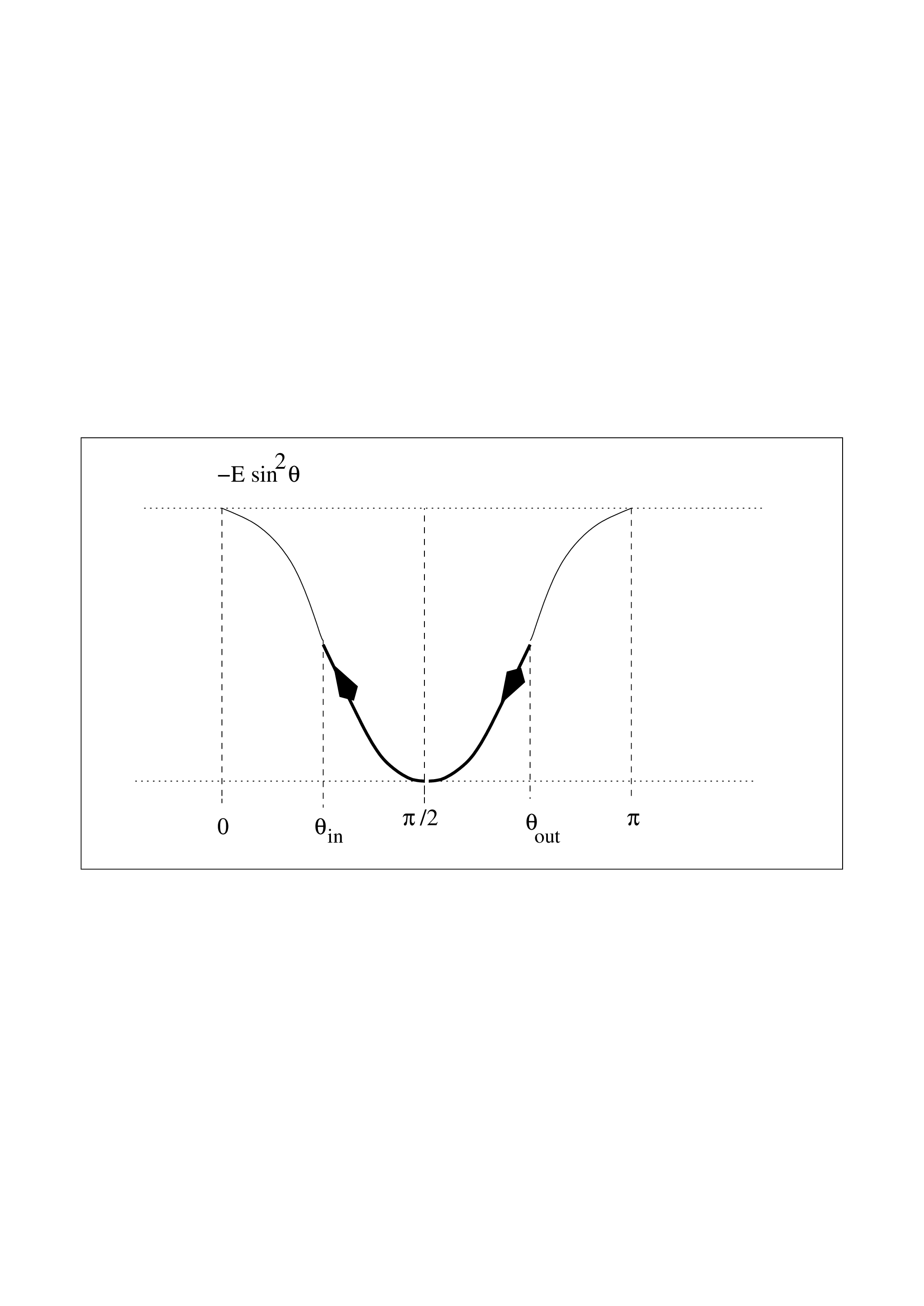}

\vspace{-2cm}

\caption{The angle evolves in a potential $\sim -E \sin^2 \theta$,
entrance and exit angle a symmetrically disposed, since the distance
along $x$ moved is the integral of $\cos \theta(t)$, which is exactly
zero if $\pi/2 - \theta_{in} = \theta_{out} - \pi/2$. }
  \label{gg}
\end{figure}

Within the wall region: 
\begin{equation}
L=  \frac{1}{4}(\dot \theta -E_w \sin \theta )^2 
\end{equation}
The  conserved quantity is:
\begin{equation}
{\cal{E}} = \dot \theta^2 - E_w^2 \sin^2 \theta
\end{equation}
and we have:
\begin{equation}
\dot \theta = \pm \sqrt{ E_w^2 \sin^2 \theta + 4{\cal{E}}}
\end{equation}
This is shown in figure \ref{gg}, and it is clear 
that the reflection law will hold, since the trajectory inside
the wall region is symmetric. 
In terms of the coordinates normal and parallel to
the wall:
\begin{eqnarray}
dx'&=&\cos \theta dt = \frac{\cos \theta}{\dot \theta} d\theta =
 \frac{\cos \theta}{\sqrt{ E_w^2 \sin^2 \theta +4 {\cal{E}}}}d\theta
\nonumber \\
dy'&=& \sin \theta dt = \frac{\sin \theta}{\dot \theta} d\theta =
\frac{\sin \theta}{\sqrt{ E_w^2 \sin^2 \theta + 4 {\cal{E}}}}
d\theta
\end{eqnarray}
We have 
$\dot x'= \cos \theta$ and (\ref{sol}) reads:
\begin{equation}
 (\dot \theta -E_w \sin \theta )_t
=  (\dot \theta -E_w \sin \theta )_{t=0} \; e^{- 2E_w x' }
\end{equation}
It is then clear that, in the limit of strong $E_w$, the time spent in
the wall region tends to zero, while the integrand in the action $e^{4E_w x' }$
is itself of order one (since the penetration $x' \sim O(1/E_w)$).
We conclude that the action during the bounce tends to zero as
 $|E_w| \rightarrow \infty$. Note that the argument is valid for quite general
bounces.


\begin{thebibliography}{99}
\bibitem{BoGa} And then the only hope is that the attractor itself is 
composed of two intertwined sets, related by some
 discrete symmetry playing for them
the role played by time-reversal for the attractor-repellor pair. See
F. Bonetto and G. Gallavotti Cmm. Math. Phys. {\bf 189}  263 
(1997).
\end{thebibliography}
\end{document}